\documentclass[a4paper, amsfonts, amssymb, amsmath,  showkeys, nofootinbib, twoside,floats]{revtex4-1}
\usepackage[english]{babel}
\usepackage[utf8]{inputenc}
\usepackage[colorinlistoftodos, color=green!40, prependcaption]{todonotes}
\usepackage{amsthm}
\usepackage{mathtools}
\usepackage{physics}
\usepackage{xcolor}
\usepackage{graphicx}
\usepackage[left=23mm,right=13mm,top=35mm,columnsep=15pt]{geometry} 
\usepackage{adjustbox}
\usepackage{placeins}
\usepackage[T1]{fontenc}
\usepackage{lipsum}
\usepackage{csquotes}
\usepackage[pdftex, pdftitle={Article}, pdfauthor={Author}]{hyperref} 

\begin{document}
\title{Droplet on a Sugar fiber}

\author{St\'ephane Dorbolo}
    \email[Correspondence email address: ]{s.dorbolo@uliege.be}
    \affiliation{F.R.S.-FNRS, UR-CESAM, PtYX lab, University of Liege}

    \author{Floriane Weyer}
      \affiliation{Haute Ecole Charlemagne, Ecole Rivageois}

    \author{Alexandre Delory}
     \affiliation{Institut Langevin, ESPCI Paris, Université PSL, CNRS, 75005 Paris, France \\ ENS Lyon, LPENSL, UMR5672, Lyon, France}

    \author{Apurav Tambe}
     \author{Zhao Pan}
    
    \affiliation{Mechanical and Mechatronics Engineering, University of Waterloo, Canada.
}

\date{\today} 



\begin{abstract} 
The motion of a water droplet on a single vertical sugar fiber is analyzed.  The fiber is positioned vertically, with the droplet placed at its pending end. If the capillary force exceeds the weight of the droplet, the droplet remains suspended at the fiber's extremity. As water dissolves, the fiber eventually breaks. Subsequently, the droplet may fall down carrying a portion of the undissolved fiber or, more interestingly, the droplet may be propelled upwards, remaining attached to the fiber. The process can then restart. A phase diagram is constructed based on the droplet's volume and the fiber's diameter. A model has been built to assess the critical droplet volume for a given fiber diameter, above which the droplet falls. Below this volume, the droplet can move upwards along the fiber.
\end{abstract}
 
\maketitle

\section{Introduction}

Droplet interaction with fiber(s) is common in nature and in daily life. That can be as simple as a rain droplet
that hits the fabric of your coat or a spider web. The equilibrium position or the motion of the droplet along a fiber is a complex situation compared to the wetting of a plane surface. Indeed, the curvature of the fiber as to be taken into account as well as the gravity or any other external force orientation. That means that the size of the droplet matters. The droplet's volume dictates the shape and the behavior of the droplet along the fiber.  Besides the fundamental challenges, the interaction droplet-fiber is relevant in numerous applications. The glass wool production process implies to use binder droplets
to ensure a good adhesion between the glass fibers \cite{PB}. Passive fibers are used in filtering processes \cite{chang_experimental_2023}, cooling process\cite{yang_bioinspired_2021} or dew harvesting \cite{bintein_kirigami_2023}. In other situations, they offer a unique configuration for studying fluid dynamics, for exemple when an immersed fiber is employed to observe the dilution of a liquid droplet in another liquid \cite{carroll_kinetics_1981,carroll_equilibrium_1986}. The natural fibers found in the biosphere also inspired scientists. Spider silk, for instance, relies on capillary forces : fine sections of silk fiber are embedded in hanging liquid droplets, creating a composite system providing the fiber's elasticity \cite{elettro_-drop_2016} - a principle that has since been adapted for dew harvesting \cite{jiang_spider-capture-silk_2023}. Fiber geometry also plays a significant role in droplet behavior. The conical shape of cactus spines allows guiding water droplets towards the base, a strategy that has led many experimental and numerical explorations of other geometrical configurations and orientations \cite{lorenceau_drops_2004,van_hulle_capillary_2021,fournier_droplet_2021,lee_multiple_2022,pan_upside-down_2016,zhang_dynamic_2023}. Additionally, it has been shown that when fibers come into contact, transport properties are enhanced \cite{leonard_droplets_2023}, and even more so when the fibers are twisted \cite{twisted}.


When the fiber is horizontal, the shape adopted by the droplet around the fiber has been demonstrated to depend on the nature of the liquid and the fiber. Yet, the volume of the liquid involved is still relevant. Moreover, besides the nature of the fiber material, its surface roughness may drastically change the droplet shape and behavior along the fiber. According to the gravity orientation with respect to the wire, to the volume of the droplet, to the contact angle and to the diameter of the fiber, the droplet may adopt either a symmetrical shape nicknamed barrel or an asymmetrical shape, named clamshell  \cite{mchale_global_2002,mullins_observation_2006,mugelesoft2011}. 

The motion of the droplet on the fiber can be triggered by  inclining the fiber. For a given droplet volume that depends again on the nature of the liquid and the fiber, the droplet starts moving downwards \cite{mullins_effect_2004,huang_equilibrium_2009,gilet_droplets_2010}. The droplet's apparent contact angles at the front and rear correspond to the advancing and receding values, respectively. Numerical simulations were performed to describe a droplet on a vertical fiber \cite{bodziony_stressful_2023}.

Besides the orientation of the gravity, the environment of the fiber matters. When exposed to a lateral wind, the droplets may move \cite{bintein_self-propelling_2019,wilson_aerodynamic_2023}. In a similar manner, vibration introduced by sound, for example, can be used to expel the droplet from the fiber \cite{poulain_sliding_2023,chang_experimental_2023}. The opposite situation, namely the impact of a droplet on a fiber, was also investigated since this problem is relevant for the capture of liquid by a network of fiber \cite{piroird_drops_2009,lorenceau_capturing_2004,lorenceau_off-centre_2009}.

Finally, the deformation of the fiber stands for an interesting degree of freedom regarding applications in textile for example. The fiber can be flexible \cite{duprat_flexible_2012}, and the presence of the liquid may change the properties of the fiber by swelling \cite{van_de_velde_spontaneous_2022}. In an analogue manner, a water droplet that sits on a sugar fiber locally changes the physical properties of the fiber. Indeed the liquid dissolves the fiber modifying the local radius of the fiber and its elasticity as the water invades the sugar structure. In this paper, we propose to study the fate of a vertical sugar fiber on the extremity of which, a droplet of water was released. Lumberjacks know that sawing the branch on which you sit results in a fatal issue. Droplets experience a different story.

\section{Experimental set up}

\begin{figure}
  \centering
  \includegraphics[width=9cm]{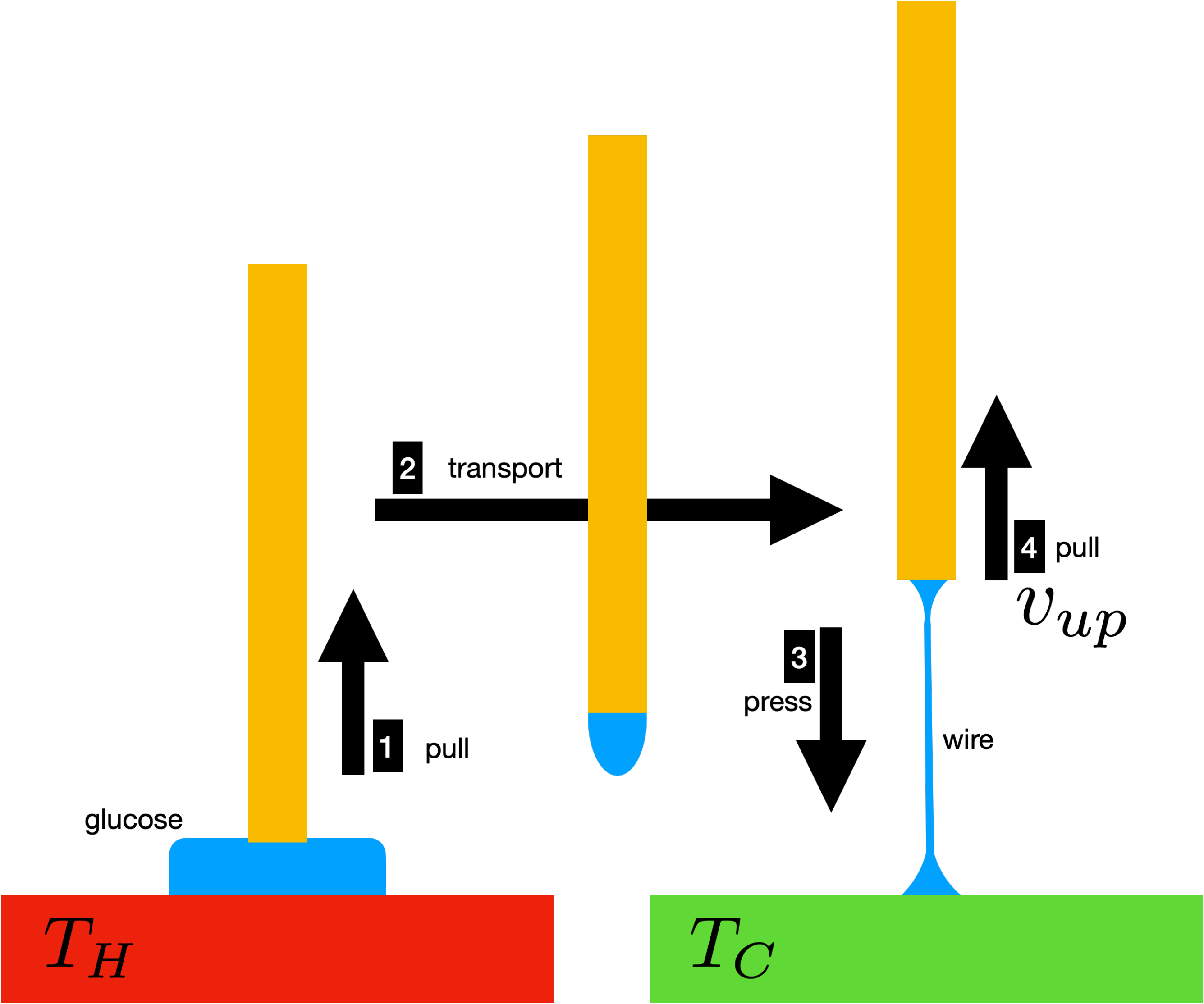}
  
  \caption{Procedure to obtain sugar fiber. (1) The solid glucose was heated at 110$^\circ$C to be melt. A rod was plunged in the liquid sugar and was pulled out of the puddle. (2) A droplet of sugar was transport towards a cold plate. (3) The droplet was squeezed in between the rod and the cold plage. (4) The rod was pulled up with a constant speed $v_{up}$ generating a fiber of sugar. }
  \label{expa}
\end{figure}

The method to fabricate the sugar fiber is similar experimental procedure can be found in ref.\cite{martincek_technology_2014}. The aim was to obtain a regular and cylindrical fiber. The sugar fiber were made from glucose (Alpha-Sigma, Aldrich Chemical Company, USA)  that was heated on a control temperature plate. The temperature was fixed to 110 $^\circ$C. At that temperature, the glucose melt without caramelizing (Fig.~\ref{expa}). While the sugar was liquid, a drop was captured using a wooden stick. The droplet was then transferred to a room temperature plastic substrate to be gently squeezed between the stick and the support. The stick was pulled up at a constant speed $v_{up}$(about 1 cm/s) extending the fiber between the plastic plate and the tip of the stick.  Sometimes, air bubbles may be trapped during the process. 

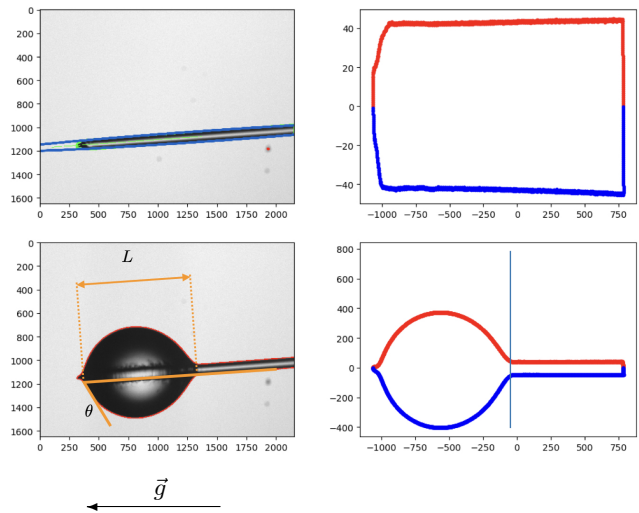
\begin{figure}
  \centering

   \begin{adjustbox}{trim=25 0 25 0, clip}
   \includegraphics[width=12cm]{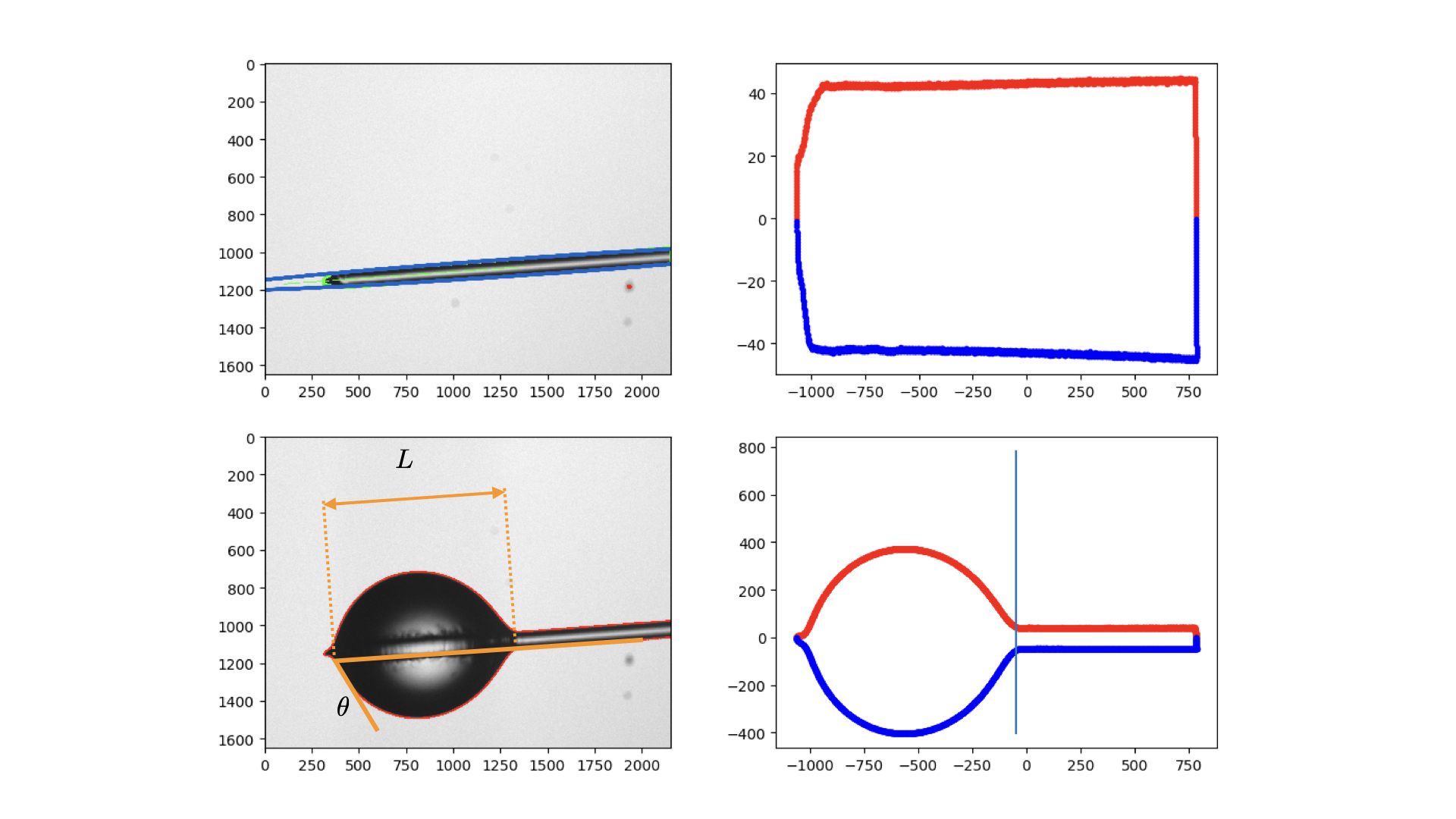}
   
\end{adjustbox}
\\

\begin{picture}(2,1)
  \put(-40,0.5){\vector(-1,0){50}} 
   \put(-65,5){$\vec{g}$}         
\end{picture} 
  \caption{Image analysis of a droplet pending at the extremity of a sugar fiber (the gravity is directed to the left hand side). (top left) and (bottom left) Images of the fiber without and with a droplet respectively. The (top right) and (bottom right) Figures correspond to the image analysis extracted from the images. Red and blue color correspond to left and right profile of the images. The vertical line in the figure (bottom right) corresponds to the determination of the contact line. In the present case, $L=$2.05 mm.}
  \label{expb}
\end{figure}

The fiber was vertically attached to an horizontal plate (Fig.\ref{expb} and \ref{diff}). The fibers did not misorient for more than a few degrees regarding the vertical direction. A water droplet was then released at the free end of the fiber in such a way that it did not wet the remaining length of the fiber. A backlight allowed obtaining highly contrasted images that were captured by a camera (JAI BM-500, Edmund Optics, USA). Three pictures were taken for each considered case (i.e. combination of a droplet of volume $V$ and a fiber of diameter $d$): the naked fiber, the fiber laden with the droplet and the outcome after dissolution of the sugar. 

In Fig.\ref{expb}, the image analysis result is shown for one case. An in-house Python code (using OpenCV package) allowed to measure the fiber diameter and to estimate the volume of the fiber covered by the droplet (first line of the four graphics in Fig.\ref{expb}). Then the code allows to extract and estimate the volume of the droplet taking into account the volume of the fiber inside the droplet. The image without droplet (top left in Fig.\ref{expb}) was first analyzed to get the shape of the fiber (top right in Fig.\ref{expb}). The gravity is oriented towards the left.  Both sides of the fiber were measured separately in red and blue, respectively. Then the image with the droplet (bottom left in Fig.\ref{expb}) allowed to extract the droplet profile (bottom right in Fig.\ref{expb}). The surface $S$ of the droplet and volume $V$ of the droplet were estimated by the rotation of the profiles (the blue and the red ones) around the axis of symmetry of the fiber. The measurements obtained for both sides were then averaged. The volume $V$ of the droplet was obtained after the subtraction of the volume of the fiber measured with the same procedure as for the droplet.

\section{Results}

The geometrical configuration of the droplet sitting at the extremity of the fiber remains unchanged during the dissolution process of the fiber which takes around minutes (according to the diameter of the fiber). This explains why three pictures are necessary and sufficient to define the dissolution dynamics from the geometrical point of view. However, some fluid motion can be observed inside the droplet during the dissolution process. The flow was evidenced because of the presence of dusts or the local density contrast due to the presence of sugar. Some movies are presented in the Supplemental Materials \cite{SM}. The observed motion of the fluid is a flow that goes downwards along the fiber and upwards along the surface of the droplet. Close to the fiber, the water is rich in sugar; the local density being larger than the rest of the droplet, the liquid locally sinks. In total, a gradient of sugar concentration builds up between the bottom and the top of the droplet. The main consequence is that the dissolution of the fiber by the droplet is not homogeneous: the fiber dissolves faster at the top of the droplet since the local concentration of sugar is lower than at the bottom. Since the fiber thins, it eventually breaks. Two possible outcomes: either the droplet detaches and falls due to gravity, or it is propelled upward and remains adhered to the fiber.

\begin{figure}
  \centering
  The droplet fell down
 
  \includegraphics[width=12cm]{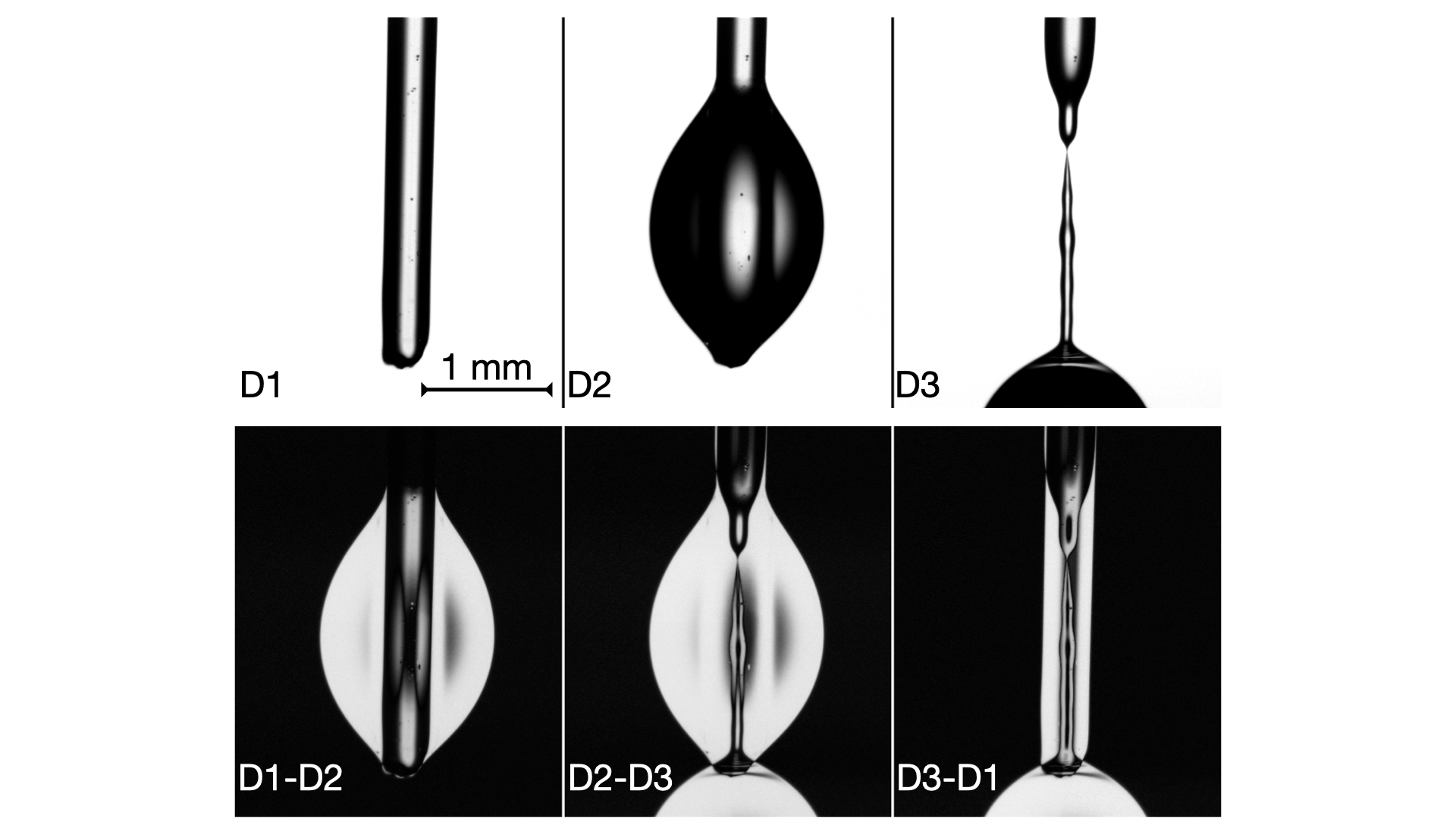}

 \vskip +5mm
 The droplet jumped upwards
 
      \includegraphics[width=12cm]{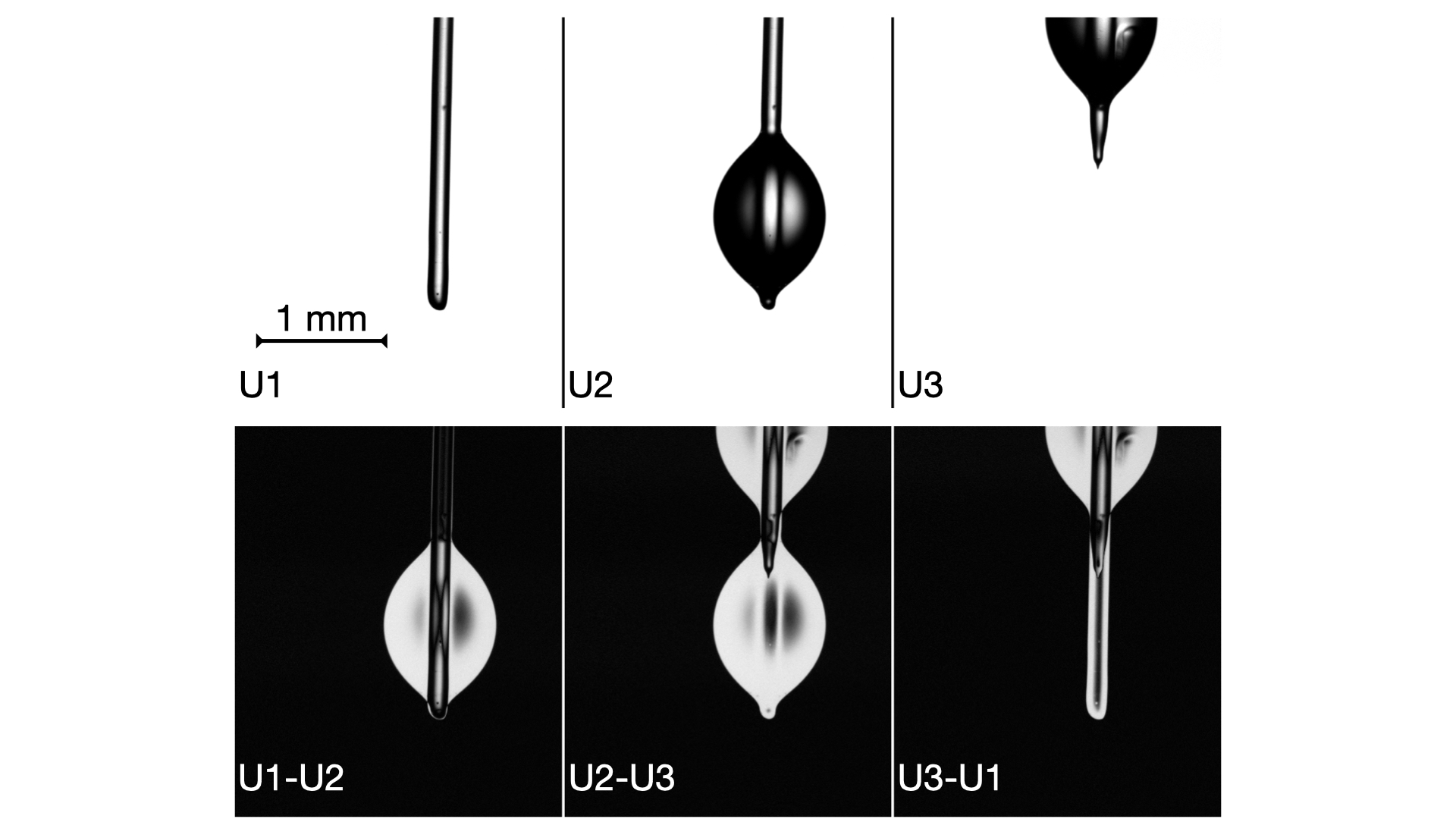}

  \caption{Presentation of the two possible outcomes of the droplet after the dissolution of the fiber either the droplet fall down (`D'-denoted picturesfor a fiber $d=387.2 \mu$m, $L=2.33$ mm, $V=1.73$ mm$^3$)  or (`U'-denoted pictures) the droplet jumped upwards ($d=162.7 \mu$m, $L=1.41$ mm, $V=0.450$ mm$^3$). The dark images result from the subtraction of two images as explicitly written, e.g. `U1-U2' means image `U1' minus image `U2'.
  }
  \label{diff}
\end{figure}

Two examples of analysed images are reported in Fig. \ref{diff}. Both scenarii are presented. The six images of the top (denoted with the letter `D') are related to the case of a droplet that fell down after having dissolved a sugar fiber of diameter $d=387.2$ $\mu$m. The six images of the bottom are for a droplet that jumped upwards on a fiber ($d=162.7$ $\mu$m) and are denoted with the letter `U'. Note that we chose a case `D' for which the fiber did not break in order to better illustrate how the fiber shape evolves due to the dissolution process.

In the first row of Fig. \ref{diff} (clear images), we can see the naked fiber on the left hand side, in the middle the pending droplet just after its release on the extremity of the fiber. Finally, on the right, the position of the droplet just after its motion at the end of the process, i.e. at the bottom of the fiber in the case `D' and at the upper part of the fiber in the case `U'. 

To better evidence the dissolution dynamics, pictures of Fig. \ref{diff} have been subtracted to each other. More specifically, the images D1-D2 and U1-U2 allow a direct visualization of the contact angles; the images D2-D3 and U2-U3 allow to evidence the motion of the droplet and finally, D3-D1 and U3-U1 show the change of shape of the fiber due to the dissolution process.

In total, 93 cases were studies for droplet volume $V$ ranging between 0.032 to 3.5 mm$^3$ for fiber diameter $d$ ranging from 125 and 542 $\mu$m. Among these 93 cases, 48 droplets jumped upwards.

\subsection{Contact angle}

The average contact angles $\theta$ at the bottom were manually measured from the images (i.e., `D1-D2' and `U1-U2') by taking the averaged values of the angles at the left and right hand side of the apparent contact line.  The contact angle at the bottom was found ranging between 20 and 60$^\circ$. On the other hand, if one zooms in on the top contact line, the contact angle is more difficult to define. Yet, we can say that the contact angle is smaller than 5$^\circ$. Indeed, the liquid seems to perfectly wet the sugar since the water is dissolving it. The observed difference between the angles measured at the bottom and at the top is due to the gravity that pulls the droplet downwards and to the presence of the edge at the extremity of the fiber that pins the contact line. 

In a first order approach, we can consider that the capillary force acting on the droplet is similar to the case of a droplet on a incline fiber. Following Furmidge idea \cite{furmidge}, this force writes $$F_c=\pi d \gamma (\cos \theta_r-\cos \theta_a),$$ where $\theta_r$ and $\theta_a$ are the receding and advanced contact angles respectively and $\gamma$ the surface tension. If one approximates the receding contact angle to zero, the calculated bottom contact angle $\theta_c$ can be found by balancing $F_c$ and the weight of the droplet as \begin{equation}\theta_c=\arccos \left (1-\frac{\rho V g}{\pi d \gamma} \right )\label{angle}\end{equation} where $\rho$ is the density of water (at first approximation). The averaged measured angle $\theta$ (i.e. the average measurement of the apparent contact angle at the right and at the left of the droplet) is plotted as a function of the calculated one, $\theta_c$, in Fig. \ref{Ltheta}a. A good correlation is found since the solid line is a fit by $a+\theta_c$ where $a$ is a fitting parameter found here to be equal to about 10$^\circ \pm 6^\circ$. Actually, regarding the main bisector (represented by the dashed line in Fig.\ref{Ltheta}a, the dispersion is rather large. However, the law in Eq.(\ref{angle}) underestimates the advancing contact angle as measured close to the extremity of the fiber. Since we wanted to work on dry fiber instead of wet fiber (see Sect.IIIE), we released the droplet directly close the extremity. If the droplets were released at any other place on the fiber, most of the droplets would have slide along the fiber. In the considered cases, the edge blocked any motion of the droplet by pinning the contact line. Consequently, the angle measured at the bottom of the droplet could be larger than the calculated advancing angle. 

\subsection{Fiber shape and droplet shape}

From `D3-D1' and `U3-U1', we can visualize the shape of the fiber inside the droplet just before the motion of the droplet. One can see the inhomogeneous dissolution of the fiber. The fiber becomes conical at the junction between the droplet and the fiber. Moreover, the fiber is the thinnest not at the top contact line but just below. This is particularly obvious and spectacular on `D2B' where the fiber thinnest dimension is of the range of three pixels (consequently smaller than 5 $\mu$m) without breaking. 

\begin{figure}
  \centering
   (a)  \includegraphics[width=8cm]{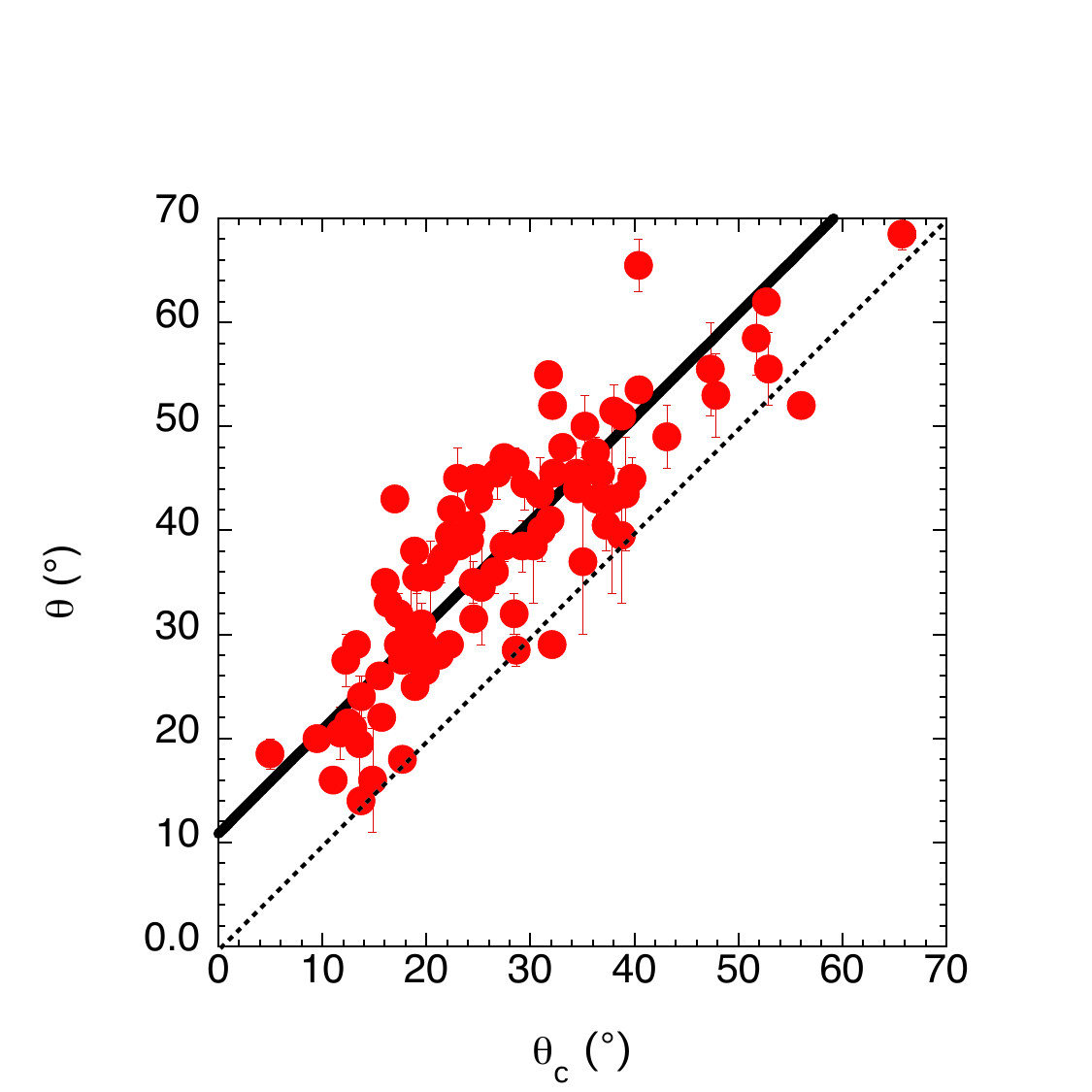}
  (b) \includegraphics[width=8cm]{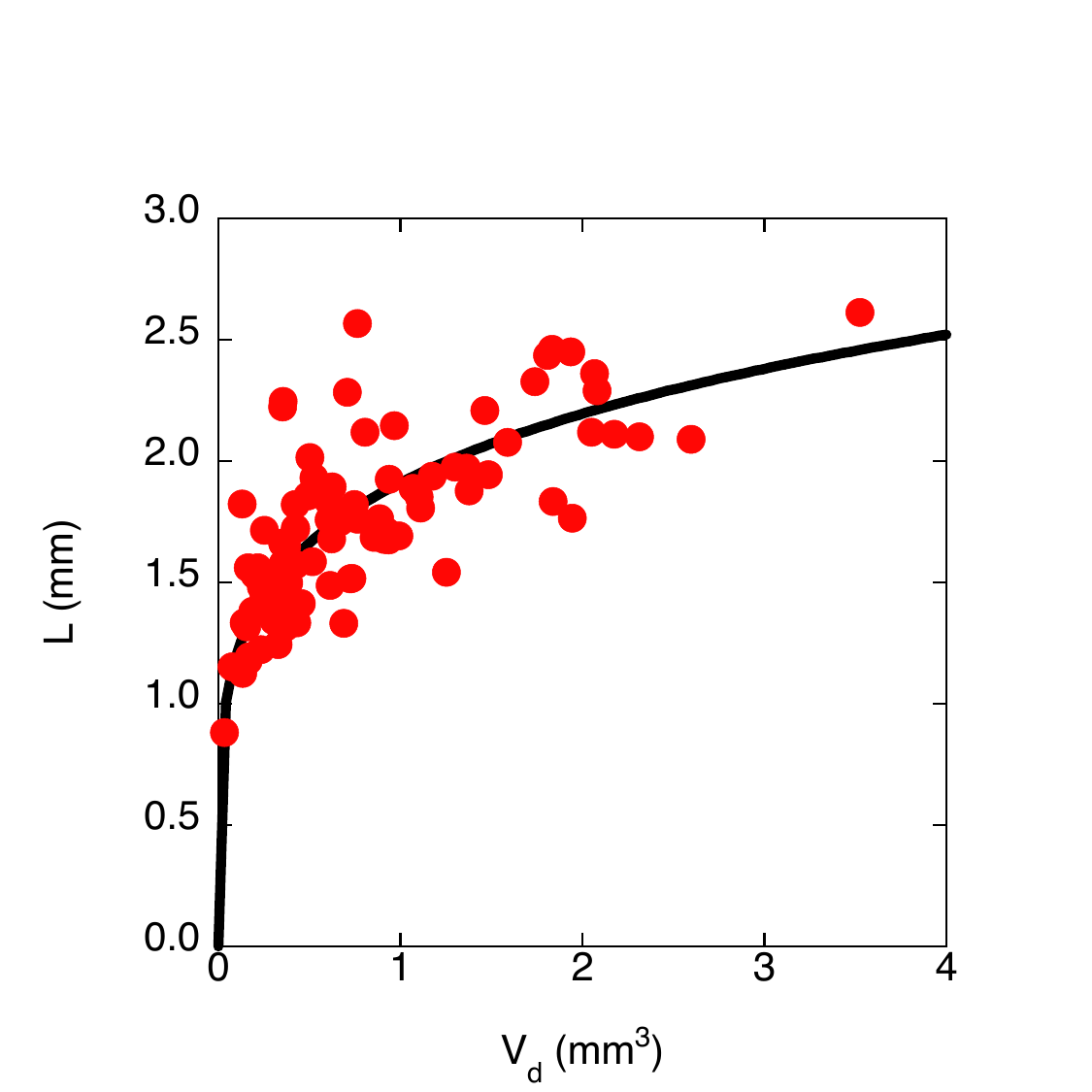}
 
  \caption{(a) Bottom contact angle $\theta$ (see text) as a function of the calculated contact angle following Eq.(1). The line is a linear fit with a slope equal to the unity. (b) Length $L$ of the droplet when sticking to the fiber as a function of the droplet volume. The curve correspond to a power law $V_d = 1.91 L^{1/5}$, that is ($V_d[\text{m}^3]=0.12 (L[\text{m}])^{1/5} $. }
  \label{Ltheta}
\end{figure}

When the droplet sits at the extremity of the fiber, regarding the geometry, three parameters can be measured (see Section II): the volume of the droplet $V$, the extension $L$ of the droplet along the fiber, and the diameter $d$ of the fiber. For a given volume $V$, the droplet spreads on a distance $L$ along the fiber and makes a contact angle $\theta$ at the bottom of the fiber (measured manually see above).  The vertical extension $L$ is reported as a function of the volume $V_d$ in Fig. \ref{Ltheta}b. The solid curve corresponds to a fit by a phenomenological power law $L = 1.91 V_d^{1/5}$. The larger the volume is, the longer the droplet spreads.


\subsection{Phase diagram}

The ninety-three droplets of volume $V$ siting of fibers of diameter $d$ have been considered to build up a phase diagram {\it Jumped or Fell}. For each couple $V-d$, the outcome of the dissolution was recorded, namely if the droplet fell down or jumped upwards on the fiber. The result is plotted in Fig. \ref{phas_dia} where the data point is a blue square when the droplet jumped and a red bullet when the droplet fell down. The blue squares and the red bullets are rather mixed for volumes below 0.7 mm$^3$. On the other hand, the bullets are mainly located above this limit. This suggests that a criterion exists to predict whether the droplet would fall down. In theory, this also corresponds to the criterion whether the droplet should jump upwards. But, in practice, failure remains possible: a droplet predicted to jump may instead be observed to fall. The reverse situation cannot occur: a droplet predicted to fall down cannot jump {\it by accident}.

\subsection{Model}
To start with, we can evaluate the criterion for the volume above which the droplet must just slide and fall.  Since, the droplet is attached to the fiber, a simple condition for the attachment is that the maximum weight $W^\star$ of the droplet is equal to the maximum capillary force $F_c=\pi d \gamma$. This condition simply reads 
\begin{equation}
    W<\pi d \gamma
    \label{weight}
\end{equation} where $\gamma$ is the surface tension of the water (72 mN/m). The force is called maximum because Eq.(\ref{weight}) considers that the wetting is total which is roughly the case since we observe a small contact angle at the upper contact line. 

\begin{figure}
  \centering
  \includegraphics[width=10cm]{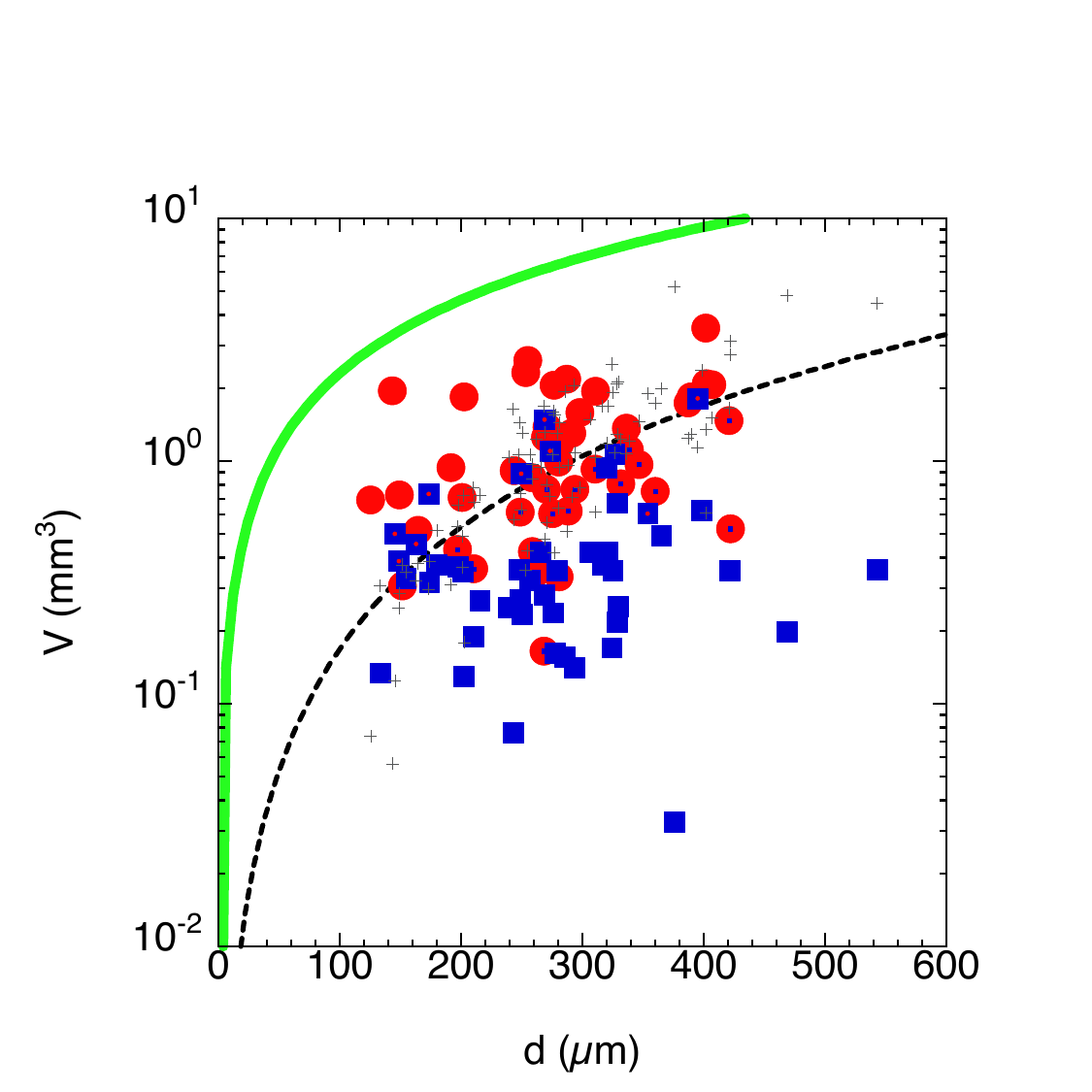}
  \caption{Phase diagram for the climbing (Blue squares) and falling droplets (Red bullets) as a function of their volume $V$ and the diameter of the fiber $d$. Below the continuous green curve, the droplet can maintain on the fiber (Eq. \ref{weight}). The gray crosses $V^\star(d)$ indicate the threshold volume below which the droplet is supposed to go upwards for each couple $V-d$; they are calculated using Eq.(\ref{scaling}). When the data point does not follow the model, an additional symbol is added in the center of the main symbol. The dashed curve is the power law $V^\star \propto d^{5/3}$ Eq.(6). }
  \label{phas_dia}
\end{figure}

As the weight is given by $\rho V g$ where $\rho$ is the density of water, we can deduce the maximum volume $V_{max}$ that a fiber of diameter $d$ can hold, i.e. $V_{max}=\pi d \gamma/(\rho g)$. This theoretical upper bound for the volume is reported as the continuous green curve in Fig. \ref{phas_dia}. All the data points are, of course, located below this limit since the reported data concerned only cases when the droplet falls after having dissolved the fiber.

We observed that the condition for dropping or jumping upwards to remain attached is related to the diameter of the fiber and the volume of the droplet. When the droplet stands on the fiber during the dissolution process, the capillary forces apply at the top and at the bottom contact angle of the droplet with the fiber. These forces tend to minimize the surface of liquid. The wetting process tends to spread the droplet along the fiber which is possible in the upward direction and not at the bottom since the contact line is pinned at the extremity of the fiber. On the whole, the piece of fiber located inside the droplet is longitudinally compressed. When the fiber breaks, the bottom capillary force $F=\pi d \cos \theta$ ($\theta$ is the contact angle at the bottom of the droplet) is not balanced any longer by the compression in the fiber. This force induces a vertical acceleration to the droplet during a typical time that is given by the capillary time $\tau_c=\sqrt{m/\gamma}$, where $m$ is the mass of the droplet, $m=\rho V$.  The maximum vertical speed $v_{jump}$ can be evaluated after a time $t=\tau_c$
\begin{equation}
v_{jump}=\frac{F}{m} \tau_c=\pi d \cos \theta \sqrt{\frac{\gamma}{\rho V}}
\end{equation}
To remain attached, the droplet must jump a height at least equal to its length $L$. As the fiber is broken, the droplet is submitted to the gravity. The minimal speed $v_{min}$ to move vertically of a distance $L$ is evaluated by 
\begin{equation}
v_{min}=\sqrt{2gL}
\end{equation} A basic criterion to remain attached is obtained by comparing $v_{jump}$ and $v_{min}$. This equality allows defining the critical volume $V^\star$ below which the droplet may jump and remain attached to the fiber,  namely
\begin{equation}
V^\star (d) = \frac{\pi^2 d^2 \gamma \cos^2 \theta}{2\rho g L}.
\label{scaling}
\end{equation} Taking the measured $\theta$ and the measured $L$, we can compute the threshold $V_c(d,\theta,L)$ for each couple of considered $V-d$. They are reported in Fig.(\ref{phas_dia}) by gray crosses. Consequently, we can evaluate whether the droplet was supposed to go upwards or to fall according to the model. If the data point is above the corresponding cross, the droplet was predicted to fall. If the data is below, the droplet was predicted to jump upwards. In case of mismatch between the prediction and the observation, the symbol was surcharged with the symbol of the prediction, namely a blue square if the droplet was supposed to go up and a red circle if the droplet was supposed to fall down. We can observe that the model fails for 6 cases (out of 48 cases for which the droplet was observed to go upwards) for which the droplet was supposed to fall. On the other hand, the model fails for 20 cases (out of 45 observations of falling) for which the droplet was supposed to jump upwards. We can conclude that the criterion can predict the fate of the droplet knowing the volume of the droplet, the contact angle and the length $L$.

\begin{figure}
  \centering
\includegraphics[width=12cm]{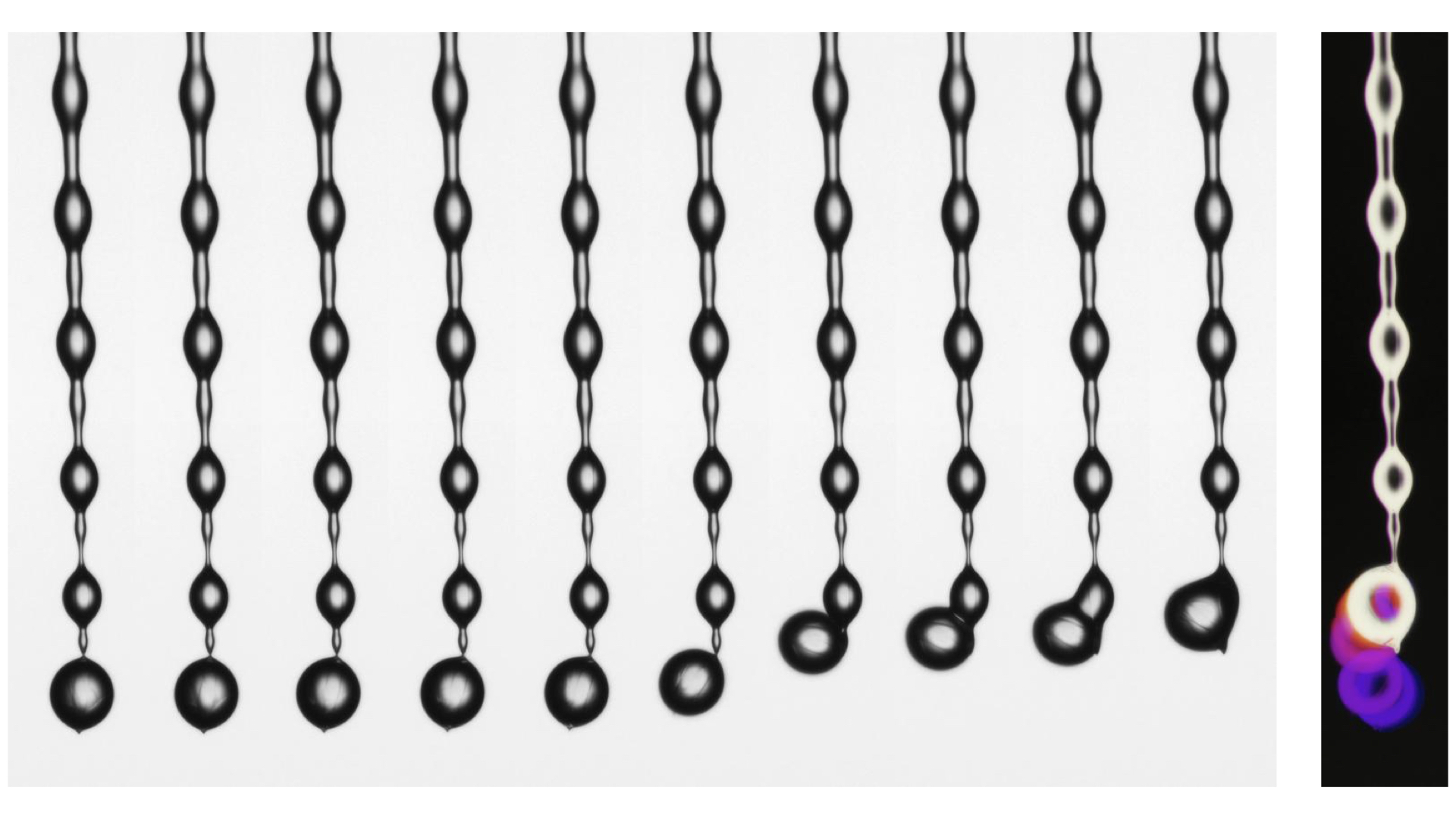}
  \caption{Successive snapshots of a pre-wet sugar fiber. The time delay between two images is 85 ms. On the right, the image is obtained by summing the successive snapshots.}
  \label{rota}
\end{figure}

Even if a model that predicts the length of the droplet is still needed, we can try to extract an empirical law for the criterion Eq.(\ref{scaling}) by introducing the phenomenological law for the extension $L = A V^{1/5}$ where $A=0.12$ in SI. If we neglect the variation of the contact angle with the volume by taking a fixed average angle $\bar \theta=38.15^\circ$, we have that \begin{equation} V^\star = 7.8 \ 10^{-5} d^{5/3}.
\label{emp}\end{equation} We plotted Eq.(6) and obtained the black dashed curve in Fig.(\ref{phas_dia}).  Since this criteria is partially based on Eq. (\ref{scaling}), it is not surprising that the black dashed curve separates the blue squares and the red circles well. In addition, the mixed symbols (red dot in blue squares or blue square dot in a red circles) are all located around the phenomenological curve. We may treat them as transit cases due to numerous influences like inhomogeneities in the fiber, vibrations,..  These observations further validate Eq. (\ref{scaling}) as a good model. Yet, the empiric scaling model Eq.(\ref{emp}) allowed to provide a rapid criterion for the attachment of the droplet after dissolution.

\subsection{Wet sugar fiber}
The situation is rather different when the sugar fiber is beforehand wet by the passage of the droplet. A large droplet was released close to the attachment point of the fiber. The droplets slid and wet the fiber before detaching when it reached the lower extremity of the fiber. After the passage of the droplet, a film of water was released along the fiber. The dissolution started processing and the liquid film made of water and dissolved sugar started to become instable along a Rayleigh-Plateau instability process. In Fig. \ref{rota}, a time lapse of such a wet fiber is shown. The time delay between two successive images is 85 ms. Since the deposition of liquid is rather minimal, one expects to observe the jump of the droplet located at the extremity of the fiber. Sometimes the droplets fall down, sometimes the droplet jumps, sometimes the fiber collapses. However, we observed (several times) an additional interesting behavior, the fiber can bend inside the droplet under the action of the surface tension that compress the fiber inside the droplet. This bending may result in the rotation of the fiber and a global upwards motion of the droplet (see \cite{SM}). Let us note that the pre-wetting is not necessary if the moisture of the air is sufficient, water adsorbs on the fiber before dissolving it according to a same scenario.

\section{Conclusion}
A droplet of water pending at the extremity of a vertical sugar fiber has a contact line located along the edge of the fiber and another located on the fiber. Surface tension compresses the fiber located inside the droplet while the sugar is dissolving. The dissolution is not homogeneous due to the induced gradient of sugar concentration in the droplet under the action of the gravity. The fiber is the thinnest at the upper part of the droplet until the fiber breaks. A criterion that account on the fiber diameter, the contact angle of the droplet, the extension of droplet and the volume of the droplet has been defined and allows to predict either the droplet falls or jumps. If this criterion is not satisfied, the droplet may manage to remain attached to the fiber to start the dissolution process again. Eventually, the droplet liquid reaches a large concentration of sugar. The consequence is the increase of the density that will result in the falling of the droplet. Other mechanisms like droplet rotation and fiber bending were observed in pre-wet fiber after the development of a Rayleigh-Plateau instability.
\section{Acknoledgements}
SD is a F.R.S.-FNRS Senior Research Associate. Part of this work was financed by UR-CESAM.

\bibliography{barbapapa}

\end{document}